\documentclass[twocolumn,showpacs,preprintnumbers,superscriptaddress,amsmath,amssymb]{revtex4}%
\usepackage{amsfonts}
\usepackage{amsmath}
\usepackage{amssymb}
\usepackage{graphicx}
\usepackage{footmisc}
\begin{document}
%\title{Quantum potential plays the role of scalar field: \\
%Inflation of small true vacuum bubble by quantization of Einstein-Hilbert action}
\title{Inflation of small true vacuum bubble by quantization of Einstein$-$Hilbert action}
\author{Dongshan He}
\affiliation{State Key Laboratory of Magnetic Resonances and Atomic and Molecular Physics,
Wuhan Institute of Physics and Mathematics, Chinese Academy of Sciences, Wuhan
430071, China}
\affiliation{Graduate University of the Chinese Academy of Sciences, Beijing 100049, China}
\author{Qing-yu Cai}
\thanks{Corresponding author. Electronic address: qycai@wipm.ac.cn}
\affiliation{State Key Laboratory of Magnetic Resonances and Atomic and Molecular Physics,
Wuhan Institute of Physics and Mathematics, Chinese Academy of Sciences, Wuhan
430071, China}

\begin{abstract}

We study the quantization of the Einstein$-$Hilbert action for a small true vacuum bubble without
matter or scalar field.
The quantization of action induces an extra term of potential called quantum potential in
Hamilton$-$Jacobi equation, which gives expanding solutions, including the exponential expansion
solutions of the scalar factor $a$ for the bubble.
We show that exponential expansion of the bubble continues with a short period,
no matter whether the bubble is closed, flat, or open.
The exponential expansion
ends spontaneously when the bubble becomes large, {\it that is}, the scalar factor $a$ of the
bubble approaches a Planck length $l_p$. We show that it is the quantum potential of the small
true vacuum bubble that plays the role of the scalar field potential suggested
in the slow-roll inflation model.
With the picture of quantum tunneling, we calculate particle creation rate during inflation,
which shows that particles created by inflation have the capability of reheating the universe.

\end{abstract}

\pacs{98.80.Qc, 98.80.Cq}
\maketitle

\section{Introduction}

The inflationary cosmology model by Starobinsky~\cite{aas79,aas80}
and Guth~\cite{ahg81} presents a way to resolve cosmological
puzzles of the flatness, horizon and the primordial monopole.
The recent detection of B modes in the polarization of the
cosmic microwave background by BICEP2 \cite{par14}
gives a solid evidence for inflationary theory of cosmology.
In Guth's original work, inflation was regarded as a delayed
first-order phase transition from the supercooled false vacuum
to the lower energy true vacuum. It was soon realized that such
a cosmological model has a serious problem called the graceful
exit problem. Soon after that, slow-roll inflation model
was suggested to overcome this problem~\cite{al821,al822,as82}.

The slow-roll model suggested by Linde~\cite{al821,al822} and
Albrecht and Steinhardt~\cite{as82} is based on the symmetry breaking
mechanism called Coleman$-$Weinberg mechanism~\cite{cw73} that allows
the phase transition to occur by forming bubbles, while the potential
barrier at low temperature is very small.
The essence of the slow-roll model is the assumption of the existence of
a scalar field $\varphi$, or called {\it inflaton}, that makes the
value of the potential $V(\varphi)$ be very large but quite flat at
the beginning. With the scalar field rolling very slowly down the
potential, the bubble experiences a nearly exponential expansion before
the field changes very much. However, scientists do not know what the
scalar field exactly is until now. A possible candidate is the Higgs
field, while the energy Higgs boson is so far from that of inflaton.

It is widely believed that true vacuum without matter or scalar
field cannot expand, at least it has no inflationary solution to
create the universe~\cite{bd12}.
That is the reason why scientists have to assume the existence of scalar
field in inflation theory.
Recently, with the de Broglie$-$Bohm quantum trajectory theory and
Wheeler$-$DeWitt equation (WDWE),
we have proven there are exponential expansion solutions of scalar
factor $a$ for a small true vacuum bubble when the operator ordering
factor takes a specific value $p=-2$ (or 4 for equivalence),
which shows the possibility of spontaneous creation of
the universe from nothing, in principle~\cite{our14}.

In this paper, we extend our previous study on the inflation for a
small true vacuum bubble by quantizing its Einstein$-$Hilbert action.
When the action of the small true vacuum bubble
is quantized with de Broglie$-$Bohm quantum trajectory method,
it induces an extra term, usually called quantum
potential, in the Hamilton$-$Jacobi equation. The quantization of the
action for a small true vacuum bubble can give an exponential
expansion solution of the scalar factor $a$ of the bubble with specific
ordering factor $p$.
We show it is the quantum potential that provides the power for inflation,
so that the assumption of the existence of scalar field $\varphi$ in the slow-roll
model is not necessary.
Numerical solutions show that the Hubble parameter is almost a constant as
$H(t)\sim1/t_{p}$ when the universe is very small
($a \lesssim l_p$). The value
of Hubble parameter decreases rapidly when the universe becomes large
($a > l_p$), and thus the inflation ends.
Quantum tunneling method is applied to calculating particle creation rate
during inflation, which shows particles created by inflation have the
capability of reheating the universe.

\section{WDWE for a true vacuum bubble}

Heisenberg's uncertainty principle indicates that a small true vacuum bubble
can be created probabilistically in a metastable false vacuum, in principle.
In fact, it is important to study the behaviors of the small true vacuum bubble
after its formation, rather than the process of bubble formation.
The small true vacuum bubble can be described by a minisuperspace
model~\cite{npn00,npn12,apk97}
with one single parameter of the scale factor $a$ since it only has one degree
of freedom, the bubble radius. The Einstein$-$Hilbert action for the vacuum bubble
can be written as
\begin{equation}\label{action}
S=\frac{c^3}{16\pi G}\int  \mathcal{R}\sqrt{-g}d^{4}x,
\end{equation}
where $c$ is speed of light and $G$ is the gravitational constant.
The bubble may be homogeneous and isotropic since it is true vacuum bubble.
So, the metric of the bubble in the minisuperspace model is given by
\begin{equation}\label{metric}
ds^{2}=\sigma^{2}\left[-N^{2}(t)c^2dt^{2}+a^{2}(t)d\Omega_{3}^{2}\right].
\end{equation}
Here, $d\Omega_{3}^{2}=dr^{2}/(1-kr^{2})+r^{2}(d\theta^{2}+\sin^{2}\theta
d\phi^{2})$ is the metric on a unit three-sphere, $N(t)$ is an arbitrary lapse
function, and $\sigma^{2}=2/3\pi$ is a normalizing factor chosen for later
convenience. It should be noted that $r$ is dimensionless and the scale
factor $a(t)$ has length dimension~\cite{Weinberg}. From Eq.~(\ref{metric}),
we can get $\sqrt{-g}=N\sigma^{4}ca^{3}$, and the scalar
curvature is given by
\begin{equation}
\mathcal{R}=6\frac{\ddot{a}}{\sigma^{2}c^2N^{2}a}+6\frac{\dot
{a}^{2}}{\sigma^{2}c^2N^{2}a^{2}}+\frac{6k}{\sigma^{2}a^{2}}.  \label{Ricci}%
\end{equation}

Inserting Eqs.~(\ref{metric}) and (\ref{Ricci}) into Eq.~(\ref{action}%
), we can get
\begin{align*}
S  &  =\frac{6\sigma^{2}Nc^4}{16\pi G}\int\left(\frac{a^{2}\ddot{a}%
}{N^{2}c^2}+\frac{a\dot{a}^{2}}{N^{2}c^2}+ka\right)d^{4}x,\\
%&  =\frac{6\sigma^{2}Nc^4}{16\pi G}\int d^{3}x\int\left(\frac{a^{2}\overset{\cdot
%\cdot}{a}}{N^{2}c^2}+\frac{a\overset{\cdot}{a}^{2}}{N^{2}c^2}+ka\right)dt,\\
&  =\frac{6\sigma^{2}Nc^4V}{16\pi G}\int\left(\frac{a^{2}\ddot{a}
}{N^{2}c^2}+\frac{a\dot{a}^{2}}{N^{2}c^2}+ka\right)dt,\\
&  =\frac{Nc^4}{2G}\int\left(-\frac{a\dot{a}^{2}}{N^{2}c^2}+ka\right)dt.
\end{align*}
The Lagrangian of the bubble can thus be written as
\begin{equation}
\mathcal{L}=\frac{Nc^4}{2G}\left(ka-\frac{a\dot{a}^{2}}{N^{2}c^2}\right), \label{L}%
\end{equation}
where the dot denotes the derivative with respect to time, $t$, and the
momentum $p_a$ is
\[
p_{a}=\frac{\partial\mathcal{L}}{\partial\dot{a}}=-\frac{c^2a\dot{a}}{NG}.%
\]
The Hamiltonian can be expressed by Lagrangian $ \mathcal{L}$ and momentum $p_a$ in the
canonical form:
\begin{equation}
\mathcal{H=}p_{a}\dot{a}-\mathcal{L}. \nonumber
\end{equation}
Taking $N=1$, we can get the Hamiltonian
\begin{equation}
\mathcal{H=-}\frac{1}{2}\left(\frac{Gp_{a}^{2}}{c^2a}+\frac{c^4ka}{G}\right).\nonumber
\end{equation}
In quantum cosmology theory, the evolution of the universe is completely
determined by its quantum state that should satisfy the WDWE. With
$\mathcal{H}\Psi=0$ and $p_{a}^{2}=-\hbar^{2}a^{-p}\frac{\partial}{\partial
a}(a^{p}\frac{\partial}{\partial a})$, we get the WDWE for the true vacuum
bubble~\cite{bd67,swh84,av94}:
\begin{equation}
\left(\frac{\hbar^{2}}{m_{p}}\frac{1}{a^{p}}\frac{\partial}{\partial a}a^{p}
\frac{\partial}{\partial a}-\frac{E_{p}}{l_{p}^{2}}ka^{2}\right)\psi(a)=0.
\label{wdwm1}%
\end{equation}
Here, $k=1,0,-1$ are for spatially closed, flat, and open bubbles,
respectively. The factor $p$ represents the uncertainty in the choice of
operator ordering. $m_{p}$, $E_{p}$, $l_{p}$, and $t_{p}$ are Planck mass,
Planck energy, Planck length, and Planck time, respectively.

\section{Quantization of the action}

Mathematically, a complex function $\psi(a)$ in Eq.~(\ref{wdwm1})
can be rewritten as
\begin{equation}
\psi(a)=R(a)\exp(iS(a)/\hbar), \label{psi}%
\end{equation}
where $R$ and $S$ are real functions. Inserting $\psi(a)$
into Eq.~(\ref{wdwm1}) and separating the equation into real and imaginary
parts, we get two equations \cite{bd52,prh93}:
\begin{align}
\frac{\hbar}{m_{p}}\left(  S^{\prime\prime}+2\frac{R^{\prime}S^{\prime}}%
{R}+\frac{p}{a}S^{\prime}\right)   &  =0,\label{wdwm2}\\
\frac{(S^{\prime})^{2}}{m_{p}}+V+Q  &  =0. \label{wdwm3}%
\end{align}
Here, $V(a)=E_{p}ka^{2}/l_{p}^{2}$ is the classical potential of the
minisuperspace, the prime denotes derivatives with respect to $a$, and $Q(a)$
is the quantum potential, which is given by
\begin{equation}
Q(a)=-\frac{\hbar^{2}}{m_{p}}\left(\frac{R^{\prime\prime}}{R}+\frac{p}{a}%
\frac{R^{\prime}}{R}\right). \label{qp}%
\end{equation}
It is easy to verify that Eq.~(\ref{wdwm2}) is the continuity equation~\cite{our14,av88}.
Eq.~(\ref{wdwm3}) is similar to the classical Hamilton$-$Jacobi equation, supplemented
by an extra term called quantum potential $Q(a)$. $R$ and $S$ in Eq.~(\ref{wdwm3})
can be obtained conveniently from $\psi(a)$ by solving Eq.~(\ref{wdwm1}) with relations:
\begin{align}
\psi(a)&=U+iW=R(a)\exp(iS(a)/\hbar),\label{rs0}\\
R^{2}&=U^{2}+W^{2},\,\,\,S=\hbar\tan^{-1}(W/U). \label{rs}%
\end{align}

It is interesting that the Einstein$-$Hilbert action of the true vacuum bubble
in Eq.~(\ref{action}) has been quantized in Eq.~(\ref{rs}).
The quantization of the action gives an extra term $Q(a)$ in
Eq.~(\ref{wdwm3}), which determines quantum behaviors of the small true
vacuum bubble~\cite{rv14,ad14,mvj14}. It is clear that a classical true vacuum bubble cannot expand,
while, as we show below, a quantized small true vacuum bubble has expanding
solutions, including exponential expansion solutions.

By analogy with cases of non-relativistic particle physics and quantum field
theory in flat space$-$time, quantum trajectories can be obtained from the
guidance relation~\cite{lpg93,npn00},
\begin{align}
\frac{\partial\mathcal{L}}{\partial\dot{a}}  &  =\frac{-c^{2}}%
{G}a\dot{a}=\frac{\partial S}{\partial a}\,,\label{gr}\\
\dot{a}  &  =-\frac{G}{c^{2}a}\frac{\partial S}{\partial a}\,. \label{gr2}%
\end{align}
Eq.~(\ref{gr2}) is a first-order differential equation, so the 3-metric for
all values of the parameter $t$ can be obtained by integration. With
Eqs.~(\ref{wdwm3}) and (\ref{gr2}), we can get the Hubble parameter of the bubble:
\begin{equation}\label{Hubble}
H(t)=\frac{\dot{a}}{a}=\frac{G}{c^{2}}\frac{\sqrt{-m_{p}(Q+V)}}{a^{2}}.
\end{equation}
Alternatively, the Hubble parameter can also be obtained from Eqs. (\ref{rs})
and (\ref{gr2}). These two methods are equivalent.

\section{Expansion solutions of quantized true vacuum bubbles}

In this section, we briefly review how to solve the WDWE of the bubble with $k=1,-1,0$,
respectively. The quantized action and hence the evolution equations of
the scalar factor $a(t)$ of the bubble can be obtained with the wave functions of
the bubble~\cite{our14}.

\subsection{The closed bubble}

In this case $k=1$, the analytic solution of Eq.~(\ref{wdwm1}) is
\begin{equation}
\psi(a)=\left(  \frac{a}{l_{p}}\right) ^{\frac{1-p}{2}} \left[  ic_{1}I_{\nu}%
\left(\frac{a^{2}}{2l_{p}^{2}}\right)-c_{2}K_{\nu} \left(\frac{a^{2}}{2l_{p}^{2}}\right) \right] ,
\label{psi1}%
\end{equation}
where $I_{\nu}$'s are modified Bessel functions of the first kind, $K_{\nu}$'s
are the modified Bessel function of the second kind, the coefficients $c_{1}$
and$\ c_{2}$ are arbitrary constants, and $\nu=|1-p|/4$.
Generally speaking, the wave function of the bubble should be complex.
Specially, if the wave function of the universe is pure real or pure imaginary,
we have $S^{\prime}=0$ so there are no expansion solutions.
For simplicity, we set $c_{1}$ and$\ c_{2}$ as
real numbers to find the expansion solutions.

Using Eqs. (\ref{rs0}) and (\ref{rs}), we can get
\begin{equation}
S=\hbar\tan^{-1} \left[-\frac{c_{1}}{c_{2}}\frac{I_{\nu} \left(\frac{a^{2}}{2l_{p}^{2}}%
\right)}{K_{\nu} \left(\frac{a^{2}}{2l_{p}^{2}}\right)} \right] \nonumber\label{sp1}%
\end{equation}
and
\begin{equation}
R=a^{(1-p)/2}\sqrt{ \left[c_{1}I_{\nu} \left(\frac{a^{2}}{2l_{p}^{2}}\right) \right]^{2}+ \left[c_{2}K_{\nu
}\left(\frac{a^{2}}{2l_{p}^{2}}\right) \right]^{2}}\,\,.\nonumber
\end{equation}
Here, we omit the sign \textquotedblleft$\pm$" and ``$l_{p}$" in front of $R$,
since they don't affect the value of $Q(a)$ in Eq.~(\ref{qp}). For small
arguments $0<x\ll\sqrt{\nu+1}$, Bessel functions take the following asymptotic
forms:
\begin{equation}
I_{\nu}(x)\sim\frac{1}{\Gamma(\nu+1)}\left(  \frac{x}{2}\right)  ^{\nu
}\nonumber
\end{equation}
and
\begin{equation}
K_{\nu}(x)\sim\frac{\Gamma(\nu)}{2}\left(  \frac{2}{x}\right)  ^{\nu
},\,\,\,\,\nu\neq0,\nonumber
\end{equation}
where $\Gamma(z)$ is the Gamma function. It is easy to get
\begin{equation}
S(a\ll l_{p})\approx-\frac{2\hbar c_{1}}{c_{2\Gamma(\nu)\Gamma(\nu+1)}}%
\left(\frac{a^{2}}{4l_{p}^{2}}\right)^{2\nu},\,\,\,\,\,\,\nu\neq0.\nonumber
\end{equation}
Using the guidance relation (\ref{gr2}), we can get the trajectories
for any small scale factor
\[
a(t)=\left\{  \begin{aligned}
&\left[\frac{(3-4\nu)\lambda(\nu)}{3}(t+t_{0})\right]^{\frac{1}{3-4\nu}
},\,\,\,\,\,\,\nu\neq 0, \,\, \frac{3}{4}, \\
&e^{\lambda(3/4)(t+t_{0})},\,\,\,\,\,\,\,\,\,\,\,\,\,\,\,\,\,\,\,\,
\,\,\,\,\,\,\,\,\,\,\,\,\,\,\,\,\,\,\,\,\,\,\,\,\,\,\,\,\,\nu=\frac{3}
{4}, \\
\end{aligned}\right.
\]
where $\lambda(\nu)=6c_{1}/(t_{p}4^{2\nu}c_{2}\Gamma(\nu)\Gamma(\nu+1))$ has
dimension of $T^{-1}$. For the case of $p=1$ (i.e., $\nu=0$), there is no
expansion solution for the WDWE no matter whether the bubble is closed, flat,
or open.

It is clear that only the ordering factor takes the value $p=-2$ (or $p=4$ for
equivalence), i.e., $\nu=3/4$, has the scale factor $a(t)$ an exponential
behavior. In this case, the quantum potential of the small true vacuum bubble is
\begin{align}
Q(a\rightarrow0)=-\frac{E_{p}}{l_{p}^{2}}\left( a^{2}+\frac{\lambda(3/4)^{2}%
}{c^{2}}a^{4}\right) . \label{q1}%
\end{align}
We find that the first term in quantum potential $Q(a\rightarrow0)$ exactly
cancels the classical potential $V(a)=E_{p} a^{2}/l_{p}^{2}$. The effect of
the second term $-E_{p} \lambda(3/4)^{2} a^{4}/l_{p}^{2}$ is quite similar to
that of the scalar field potential in~\cite{hhh12} or the cosmological
constant in~\cite{dhc05} for inflation. Numerically solutions for the evolution
of Hubble parameter $H=\dot a/a$ of the closed bubble will be discussed later.

\subsection{The open bubble}

For the case $k=-1$, the analytic solution of Eq.~(\ref{wdwm1}) is found to be
\begin{equation}
\psi(a)=\left(  \frac{a}{l_{p}}\right)  ^{\frac{1-p}{2}}\left[  ic_{1}J_{\nu}\left(
\frac{a^{2}}{2l_{p}^{2}}\right)  +c_{2}Y_{\nu}\left(  \frac{a^{2}}{2l_{p}^{2}%
}\right)  \right]  ,\label{psi-1}%
\end{equation}
where $J_{\nu}$'s are Bessel functions of the first kind, and $Y_{\nu}$'s are
Bessel function of the second kind and $\nu=|1-p|/4$. With the relations in
Eqs. (\ref{rs0}) and (\ref{rs}), we can get
\begin{equation}
S=\hbar\tan^{-1}\left[  \frac{c_{1}}{c_{2}}\frac{J_{\nu}(\frac{a^{2}}%
{2l_{p}^{2}})}{Y_{\nu}(\frac{a^{2}}{2l_{p}^{2}})}\right]  ,\nonumber
\end{equation}
and
\begin{equation}
R=a^{(1-p)/2}\sqrt{\left[  c_{1}J_{\nu}\left(  \frac{a^{2}}{2l_{p}^{2}%
}\right)  \right]  ^{2}+\left[  c_{2}Y_{\nu}\left(  \frac{a^{2}}{2l_{p}^{2}%
}\right)  \right]  ^{2}}\,\,.\nonumber
\end{equation}
For small arguments $0<x\ll\sqrt{\nu+1}$, Bessel functions take the following
asymptotic forms, $J_{\nu}(x)\sim(x/2)^{\nu}/\Gamma(\nu+1)$, and $Y_{\nu
}(x)\sim-\Gamma(\nu)2^{\nu-1}/x^{\nu}$ for $(\nu\neq0)$. So, we have
\begin{equation}
S(a\ll1)\approx-\frac{\hbar\pi c_{1}}{c_{2\Gamma(\nu)\Gamma(\nu+1)}}\left(
\frac{a^{2}}{4l_{p}^{2}}\right)  ^{2\nu},\,\,\,\,v\neq0.\nonumber
\end{equation}
and
\[
a(t)=\left\{  \begin{aligned}
& \left[\frac{(3-4\nu)\bar\lambda(\nu)}{3}(t+t_{0})\right]^{\frac{1}{3-4\nu}
}, \,\,\,\,\,\nu\neq 0, \,\, \frac{3}{4} \\
&e^{\bar\lambda(3/4)(t+t_{0})}, \,\,\,\,\,\,\,\,\,\,\,\,\,
\,\,\,\,\,\,\,\,\,\,\,\,\,\,\,\,\,\,\,\,\,\,\,\,\,\,\,\,\,\,\,\,\,\,\,
\nu=\frac{3}{4},
\end{aligned}\right.
\]
where $\bar{\lambda}(\nu)=3\pi c_{1}/(t_{p}4^{2\nu}c_{2}\Gamma(\nu)\Gamma
(\nu+1))$.

Similarly, the scale factor $a(t)$ has an exponential behavior for the
special case of $p=-2$ (or 4). In this case, the quantum potential for
the bubble can be obtained as
\begin{equation}
Q(a\rightarrow0)=\frac{E_{p}}{l_{p}^{2}}\left(  a^{2}-\frac{\bar{\lambda
}(3/4)^{2}}{c^{2}}a^{4}\right). \label{q-1}%
\end{equation}
The terms $a^{2}$ in quantum potential $Q(a\rightarrow0)$ and classical potential
$V(a)$ cancel each other exactly. Thus, it is the second term
$-E_{p}\bar{\lambda}(3/4)^{2}a^{4}/l_{p}^{2}c^{2}$ in quantum
potential $Q(a\rightarrow0)$ that causes the exponential expansion.

\subsection{The flat bubble}

For the case of $k=0$, the analytic solution of Eq.~(\ref{wdwm1}) is
\begin{equation}
\psi(a)=\frac{ic_{1}}{1-p}\left(  \frac{a}{l_{p}}\right)  ^{1-p}%
-c_{2},\label{psi0}%
\end{equation}
where $p\neq1$, and hence
\begin{align}
S &  =\tan^{-1}\left[  -\frac{c_{1}}{c_{2}(1-p)}\left(  \frac{a}{l_{p}%
}\right)  ^{1-p}\right]  ,\,\,\,\,p\neq1,\nonumber\\
R &  =\sqrt{c_{2}^{2}+\left(  \frac{c_{1}}{1-p}\frac{a^{1-p}}{l_{p}^{1-p}%
}\right)  ^{2}},\,\,\,\,p\neq1.\nonumber
\end{align}
With the guidance relation (\ref{gr2}), we can get the form of time-dependent
scalar factor $a(t)$ as
\[
a(t)=\left\{  \begin{aligned}
&\left[\frac{c_1}{c_2}(3-|1-p|)\frac{(t+t_0)}{t_p }\right]^
{\frac{1}{3-|1-p|}}, \, |1-p|\neq 0,3, \\
& e^{\frac{c_1}{c_2}\frac{(t+t_0)}{t_p}},\,\,\,\,\,\,\,\,\,
\,\,\,\,\,\,\,\,\,\,\,\,\,\,\,\,\,\,\,\,\,\,\,\,\,\,\,
\,\,\,\,\,\,\,\,\,\,\,\,\,\,\,\,\,\,\,\,\,\,\,\,\,\,\,\,
|1-p|=3.
\end{aligned}
\right.
\]

It is clear that only the ordering factor takes the value $p=-2$ (or 4), will
the small true vacuum bubble have the exponential expansion solutions.
The accompanying quantum potential for the flat bubble is
$Q(a\rightarrow0)=-E_p (c_{1}/c_{2})^{2}a^{4}/l_p^4$, while the
classical potential is $V(a)=0$ on this condition.
This definitely means that it is the quantum potential $Q(a)$ that is the
origin of exponential expansion for the small true vacuum bubble.

\section{Hubble parameter and quantum potential}

From the discussion above, we can see that both Hubble parameter and quantum
potential of the bubble depend on three parameters: the operator ordering $p$,
the boundary condition $c_1/c_2$, and the initial condition $a_0$.
In this section, we study the time-dependent evolutions of Hubble parameters
and quantum potential numerically with different $p$ and $c_1/c_2$.

\subsection{Hubble parameter with different $p$}

With the real part $R(a)$ of wave function, we can get the value of quantum potential
$Q(a)$ using Eq.~(\ref{qp}). The evolution of Hubble parameter can thus be
obtained from Eq.~(\ref{Hubble}).

\begin{figure}[ht]
\centering
\includegraphics[height=5cm]{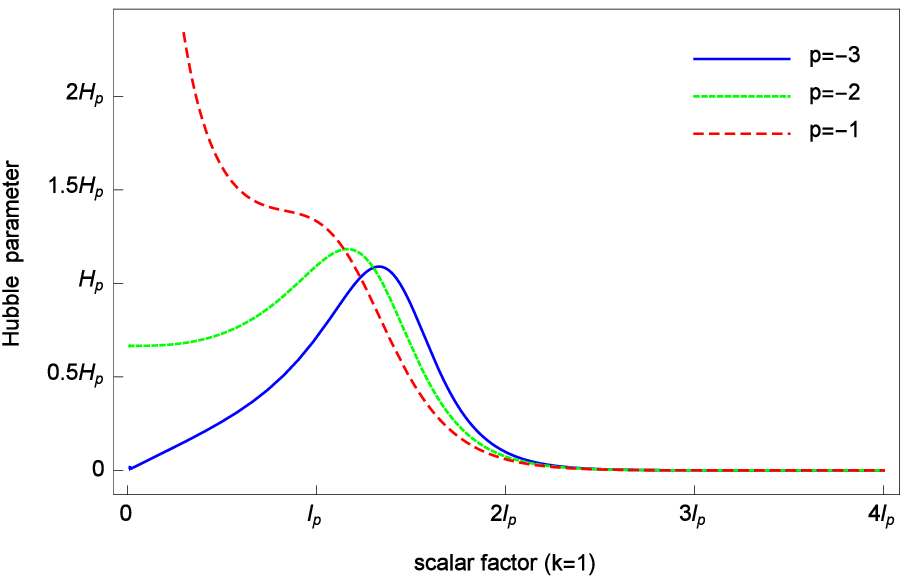}
\includegraphics[height=5cm]{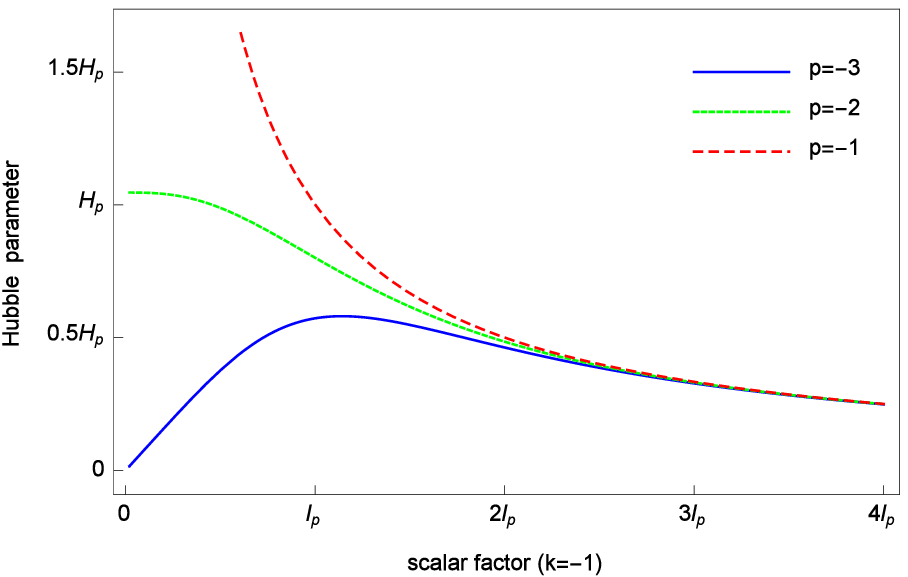}
\includegraphics[height=5cm]{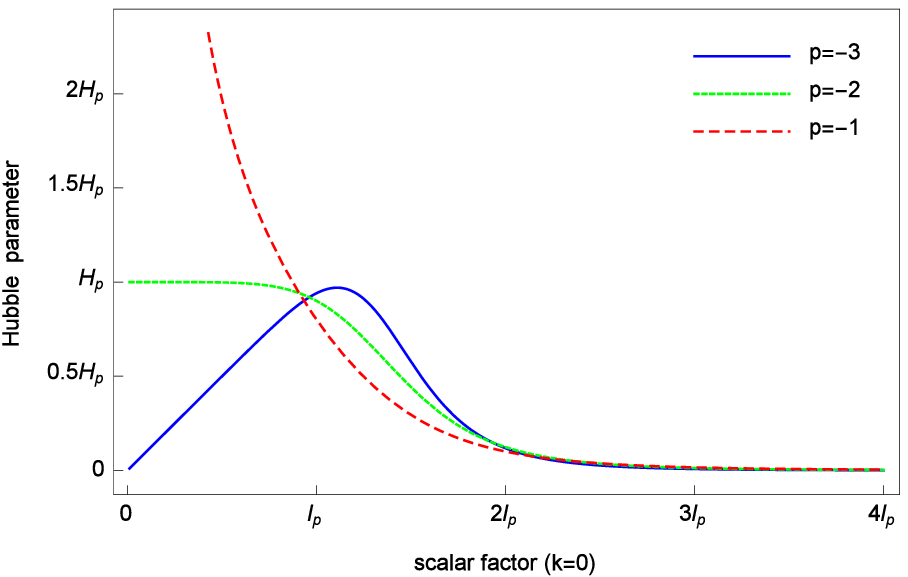}
\caption{(color online)\,
Hubble parameters for closed,open and flat bubbles,
respectively. Here, we have set $c_1/c_2=1$.}
\label{pHubble}
\end{figure}

Detailed calculations show that when the bubble is very small, that is, $a\rightarrow 0$,
Hubble parameters are divergent for $\nu < 3/4$, and it approaches zero
for $\nu>3/4$. Only $\nu$ takes value $\nu=3/4$, that is, $p=-2$ (or 4 for equivalence),
has the bubble exponential expansion solutions.
In the limit of large bubble, different operator ordering factors give the same
behavior of Hubble parameters.
Explicitly numerical solutions can be found in Fig.~\ref{pHubble}.
It is clear that the effect of the operator ordering $p$ is significant when the bubble
is small (i.e., $a\lesssim l_p$), while its effects are too small to be negligible when the bubble
becomes large (i.e., $a\gg l_p$).
In this case, we can conclude that the ordering factor $p$ represents quantum effects
of the bubble as described by Eq.~(\ref{wdwm1}).

\subsection{Quantum potential of a true vacuum bubble}

We have shown that the small true vacuum bubble expands exponentially
no matter the bubble is closed, open, or flat as long as the ordering factor takes
a specific value $p=-2$ (or 4).
As discussed previously, it is the quantum potential that provides power for inflation.
In the following, we study the evolutions of quantum potential of the bubble with the
increase of $a$. For simplicity, we set $p=-2$ and $c_1/c_2=1$ in numerical solutions.

\begin{figure}[ht]
\centering
\includegraphics[height=5cm]{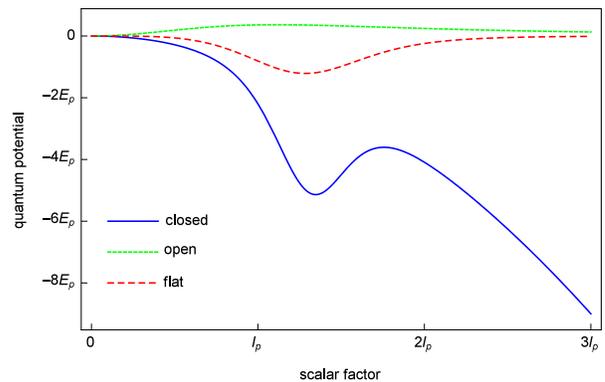}
\caption{(color online) \, The value of quantum potential for the closed,
open, and flat bubbles with the variation of $a$.}
\label{figqp}
\end{figure}

\begin{figure}[hb]
\centering
\includegraphics[height=5cm]{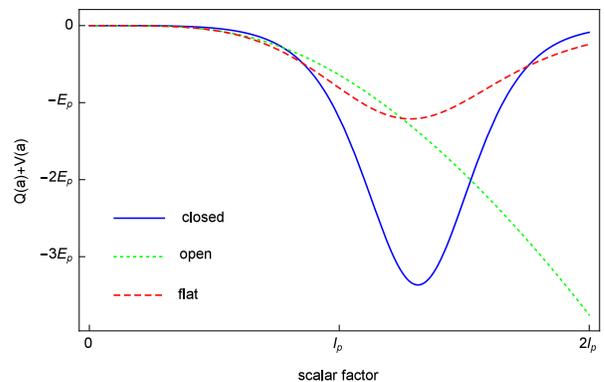}
\caption{(color online) \, The value of quantum potential plus classical potential
(that is, the value of $V(a)+Q(a)$) for closed, open, and flat bubbles versus $a$.}
\label{qplusv}
\end{figure}

In Fig.~\ref{figqp}, we find that $Q(a\gg l_p)\rightarrow0$ for the open
and flat bubbles, which indicates the quantum effects can be neglected when
the bubble becomes large enough. For the closed bubble, the asymptotic behavior
of its quantum potential is $Q(a\gg1)\sim-a^{2}$, which exactly cancels the
value of classical potential. This indicates that the quantum effect of the
closed bubble is significant no matter how large the bubble is.
According to the de Broglie$-$Bohm quantum trajectory theory, the closed bubble
should be in a steady state in the large limit.
Thus, we can conclude that the scale factor $a$ of a small true vacuum bubble
stops accelerating when the bubble becomes very large, no matter whether the
bubble is closed, open, or flat.

When the bubble is very small, that is, $a\lesssim l_p$, the sum of
quantum potential plus classical potential is directly proportional to $-a^4$,
$Q(a)+V(a)\sim -a^4$, no matter whether the bubble is closed, open, or flat.
As shown in Fig.~\ref{qplusv}, $Q(a)+V(a)$ changes very slowly when $a$ is small
($a\lesssim l_p$),
while it decreases quickly when $a$ of the bubble becomes large ($a>l_p$), which
completely satisfies the slow-roll inflation conditions~\cite{al821,al822,as82}.
Here, we can conclude that it is the quantum potential that provides the power for
the vacuum bubble inflation, which plays the role of the assumed scalar field in
the slow-roll inflation theory.

\begin{figure}[ht]
\centering
\includegraphics[height=5cm]{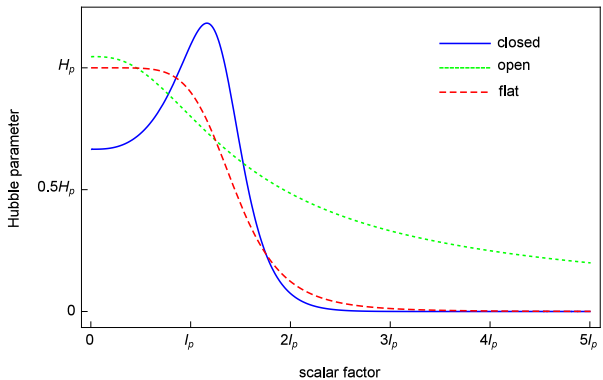}
\caption{(color online) \, The evolution of Hubble parameter with scale factor for closed,
open, and flat bubbles, respectively, with the operator ordering factor $p=-2$ and
$c1/c2=1$.}
\label{figHubble}
\end{figure}

Numerical solutions in Fig.~\ref{figHubble} show that the Hubble parameter $H$
is almost a constant when the bubble is very small, i.e., $a\ll l_p$.
For the closed or flat bubbles, Hubble parameters decrease to zero when the scalar
factor $a$ becomes large enough ($a\gg 3 l_p$). For an open large bubble,
its Hubble parameter is inverse to $a$ (i.e., $H(a)\sim 1/a$),
which means the bubble expands
with a constant velocity when the bubble is very large. Then, we can get the
conclusion again: the vacuum bubble will stop accelerating when it becomes very large,
no matter whether it is closed, flat, or open.

\subsection{Hubble parameter with different $c_1/c_2$}

\begin{figure}[ht]
\centering
\includegraphics[height=5cm]{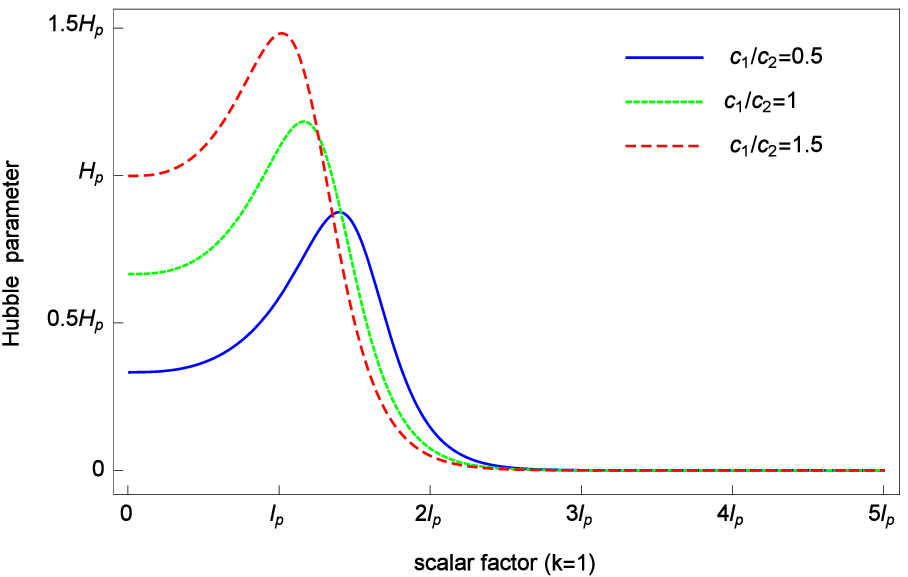}
\includegraphics[height=5cm]{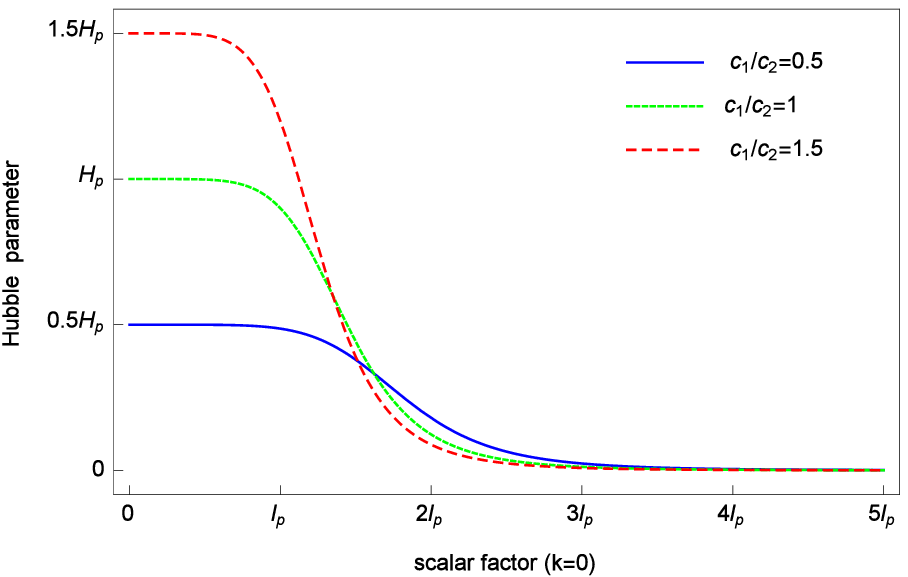}
\caption{(color online) \,
Hubble parameter for closed or flat bubbles with different $c_1 / c_2$.}
\label{cHubble}
\end{figure}

\begin{figure}[ht]
\centering
\includegraphics[height=5cm]{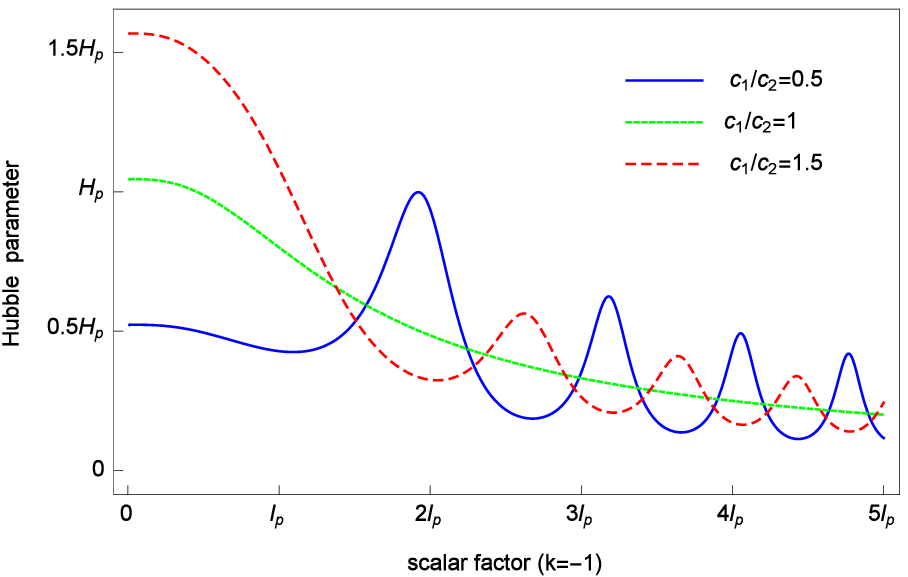}
\includegraphics[height=5cm]{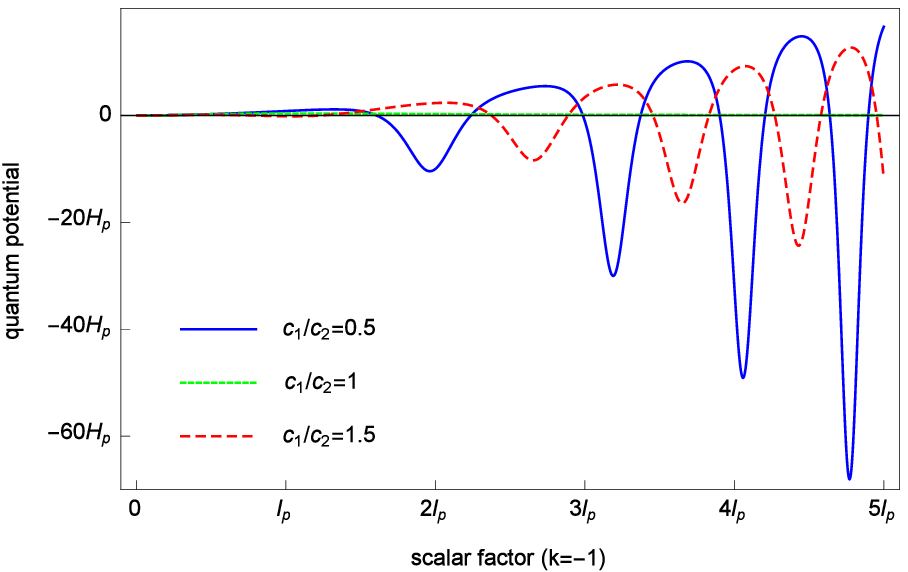}
\includegraphics[height=5cm]{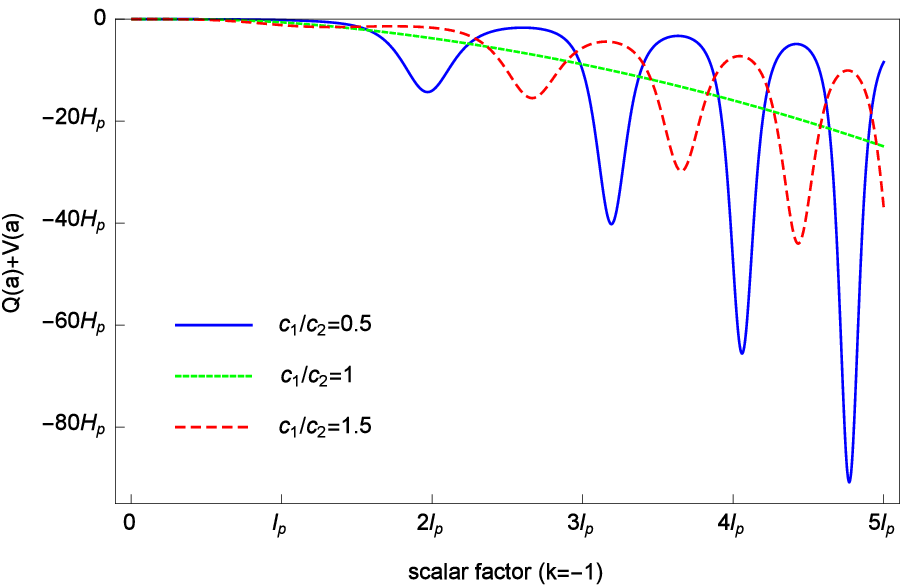}
\caption{(color online) \,
Hubble parameters and the accompanying quantum potential
for open bubbles with different $c_1 / c_2$.}
\label{Hubble:open:bubble}
\end{figure}

We study the case of inflation solutions ($p=-2$) with different values
of $c_1/c_2$.
Numerical solutions in Fig.~\ref{cHubble} show that the evolutions of
Hubble parameter have similar form with different values of $c_1 /c_2$ for
closed or flat vacuum bubble.
However, for open bubbles, Fig.~\ref{Hubble:open:bubble} shows Hubble parameters
decrease with oscillations as $a$ increases when $c_1 /c_2\ne 1$.
For a bubble with finite value of $c_1/c_2$, its Hubble parameter approaches
zero when the bubble becomes large
enough. The oscillations of
quantum potential increases while the oscillations of the companying Hubble
parameters
decrease with the increase of $a$.
This implies that the inflation will exit when the bubble becomes large enough,
no matter the bubble is closed, open, or flat.

\subsection{The e-folding number}

Let us consider how long the inflation sustains. It has been shown above
that inflation will exit when the scale factor approaches Planck length,
$a\sim l_p$.
During inflationary stage, we have $a\approx a_0 e^{H t}$, where $a_0$ is the
initial value of $a$ at $t=0$.
So, the e-folding number can be obtained as
\begin{equation}\label{e:folding:number}
N\sim\ln{\frac{l_p}{a_0}}\approx H t_{end}.
\end{equation}
The e-folding number $N$ is determined by the initial condition of the bubble,
$a_0\sim l_p e^{-N}$. The early universe has enough time ($t_{end}\approx Nt_p$)
to maintain inflation as long as $a_0$ is small enough.
In practice, the standard cosmological model requires the e-folding number $N\gtrsim 60$,
which implies that the exponential expansion of the early universe continues at least
more than 60 Planck time.

\section{Particle creation by inflation}

There is no doubt that the space and time of the early universe
will emerge by the exponential expansion of the bubble. One may
ask an important question how matter appears in the early universe.
In the scalar field inflationary model, almost all matter,
antimatter, and photons were produced by the energy of scalar field
that was released following the phase transition. However, in our
calculations, it is the quantum potential that provides the power for inflation.
In the following, we show that particles can be created by the exponential
expansion of the bubble with quantum tunneling mechanism.

The change of the spacetime metric at the end of inflation will itself create
particles due to their coupling to the spacetime curvature, which has been
discussed in many papers~\cite{lp69,asl78,ndb82,lhf87}.
In 2000, Parikh and Wilczek applied quantum tunneling method to
Hawking radiation~\cite{mkp00}. This method gives a nonthermal spectrum
which has been used to recover the lost information in Hawking
radiation~\cite{zbc09a,zczy13}.
Another method called Hamilton$-$Jacobi method~\cite{kt99,lgr11} has also
been suggested to obtain the tunneling probability.
We apply this method to inflationary universe which has a de Sitter
spacetime (the metric for a free observer who stays in an exponential
expansion spacetime) to calculate particle creation.

\subsection{The Friedmann$-$Robertson$-$Walker (FRW) spacetime}

The line element of a homogeneous and isotropic universe
can be written as
\begin{equation}
ds^{2}=-dt^{2}+a^{2}(t)d\Omega^{3}. \label{frw  metric}%
\end{equation}
The metric above is equivalent to that in Eq.~(\ref{metric}).
Here, we omit the normalizing factor $\sigma$ in Eq.~(\ref{metric}),
because the normalizing factor doesn't influence our calculation.
During the inflationary phase, the scale factor takes the form
$a(t)=e^{Ht}$, and $H=\overset{.}{a(t)}/a(t)$ is Hubble parameter.
For simplicity, we set $\hbar=c=k_{B}=1$ hereafter.

The frame of an observer at some spacetime point in the de Sitter phase
is described by the static coordinates. The static de Sitter coordinates
$(\widetilde{t},\widetilde{r},\theta,\varphi)$ are given in terms of the FRW
coordinates $(t,r,\theta,\varphi)$ by~\cite{rhb85}
\begin{align}
\widetilde{r}  &  =e^{Ht}r,   \label{tr}\\
\widetilde{t}  &  =-\frac{1}{2H}\ln\left[  e^{-2Ht}-(rH)^{2}\right]
. \label{tr2}
\end{align}
These relations are valid in the region $\widetilde{r}<1/H$.
From Eqs.~(\ref{tr}) and (\ref{tr2}), we can get
\begin{align}
dt  &  =d\widetilde{t}-\frac{H\widetilde{r}}{1-H^{2}\widetilde{r}^{2}%
}d\widetilde{r},\label{dtdr}\\
dr  &  =-\frac{e^{-H\widetilde{r}}H\widetilde{r}}{\sqrt{1-H^{2}\widetilde
{r}^{2}}}d\widetilde{t}+\frac{e^{-H\widetilde{r}}}{\left(  1-H^{2}%
\widetilde{r}^{2}\right)  ^{3/2}}d\widetilde{r}.\label{dtdr2}
\end{align}
Inserting Eqs.~(\ref{tr})$-$(\ref{dtdr2}) into FRW line element (\ref{frw metric}),
we can get the static de Sitter metric:
\begin{equation}
ds^{2}=-(1-H^{2}\widetilde{r}^{2})d\widetilde{t}^{2}+(1-H^{2}\widetilde{r}%
^{2})^{-1}d\widetilde{r}^{2}+\widetilde{r}^{2}d\Omega^{2}. \label{metric2}%
\end{equation}
Here, we have set $k=0$ for simplicity (it has been discussed that $k$ is unimportant
for particle creation during inflation~\cite{sd12,lgr11}).
Taking the substitution $H^{2}\rightarrow2M/\widetilde{r}^{3}$, we can find
that this metric is similar to the usual Schwarzschild metric,
and the horizon is located at $\widetilde{r}_{H}=1/H$.

The static metric has a limitation that it is only valid up to the
horizon, since there is a singularity at $\widetilde{r}_{H}=1/H$.
It is well known
that Painlev\'{e} metric can describe across-horizon phenomena of
a black hole~\cite{mkp00,lgr11}. Then, we translate the de Sitter
metric to Painlev\'{e}-type metric via the transformations
\[
d\widetilde{t}=dT-\frac{H\widetilde{r}}{1-H^{2}\widetilde{r}^{2}%
}d\widetilde{r}.
\]
With this choice, the metric (\ref{metric2}) reads%
\begin{equation}
ds^{2}=-(1-H^{2}\widetilde{r}^{2})dT^{2}-2H\widetilde{r}dT%
d\widetilde{r}+d\widetilde{r}^{2}+\widetilde{r}^{2}d\Omega^{2}.
\label{metric3}%
\end{equation}
At a fixed time, the spatial geometry described by (\ref{metric3}) is Euclidean,
while at any fixed radius, the boundary geometry is the same as that in
(\ref{metric2}). The metric is no more singular at the horizon
$\widetilde{r}_{H}$. Furthermore, the spacetime is stationary,
but no more static. The $T$ coordinate is nothing more than
the proper time along a radial geodesic worldline,
such as a free-falling observer.

\subsection{Tunneling across the cosmological horizon}

The great utility for a coordinate system which is well behaved at the
horizon is that one can study across-horizon physics. In this section, we
study a scalar field placed in a background spacetime. Physically, these
fields come from vacuum fluctuations that permeate the spacetime given
by the metric.
The minimally coupled Klein$-$Gordon equation for a scalar field $\phi$ of mass
$m$ in curved spacetime $g_{\mu\nu}$ has the form~\cite{ndb82}:
\begin{equation}
\left[  \frac{1}{\sqrt{-g}}\partial_{\mu}\left(  \sqrt{-g}g^{\mu\nu}%
\partial_{\nu}\right)  -\frac{m^{2}c^{2}}{\hbar^{2}}\right]  \phi=0.
\label{klein}%
\end{equation}
Inserting the scalar field in terms of a phase factor as $\phi=\phi
_{0}e^{iS(t_{p},\widetilde{r})/\hbar}$ into Eq.~(\ref{klein}), and taking
the limitation $\hbar\rightarrow0$, we can get the Hamilton$-$Jacobi equation
for the action $S$ of the field $\phi$ in the gravitational
background \cite{adg10}:
\begin{equation}
g^{\mu\nu}\left(  \partial_{\mu}S\right)  \left(  \partial_{\nu}S\right)
+m^{2}=0 . \label{h-j}%
\end{equation}
For stationary spacetime, the action $S$ can be split into two part, the time
part and space part, $S(T,\widetilde{r})=ET+S_{0}(\widetilde{r})$. In the
Wentzel$-$Kramers$-$Brillouin limit,
the probability of tunneling is related to the imaginary part of the action
for the classically forbidden trajectory~\cite{mkp00}:
\begin{equation}
\Gamma\sim e^{-2\operatorname{Im}S}. \label{gamma}%
\end{equation}

Applying the stationary Painlev\'{e} metric in Eq.~(\ref{metric3}) to the
Hamilton$-$Jacobi in Eq.~(\ref{h-j}), we can get
\[
-E^{2}-2EH\widetilde{r}\partial_{\widetilde{r}}S+(1-H^{2}\widetilde{r}%
^{2})\left(  \partial_{\widetilde{r}}S\right)  ^{2}+m^{2}=0.
\]
In this case, the action $S$ can be obtained as
\begin{equation}
S=\int\frac{EH\widetilde{r}}{1-H^{2}\widetilde{r}^{2}}d\widetilde{r}%
\pm\int\frac{\sqrt{E^{2}-m^{2}(1-H^{2}\widetilde{r}^{2})}}{1-H^{2}%
\widetilde{r}^{2}}d\widetilde{r}. \label{s0}%
\end{equation}
Here, the positive and negative signs indicate ingoing and outgoing particles,
respectively. It should be pointed out that the energy of ingoing particle
is positive and the energy of outgoing particle is negative, which are different
from those in Hawking radiation as tunneling.
The contour integral includes a
singularity at $\widetilde{r}=1/H$ and it has to be made by going around the
pole at singularity. In this way, we can obtain the imaginary part of
the ingoing particle as
\[
\operatorname{Im}S=\frac{\pi E}{H}.
\]
When we consider outgoing particles,
a minus sign should be added to the
first term in the right of Eq.~(\ref{s0})~\cite{zbc09b}.
Similarly, we can get $\operatorname{Im}S=\pi E/H$
for the outgoing particles.
Finally, we can obtain the tunneling probability for two channels as
\begin{equation}
\Gamma\sim e^{-\frac{2\pi E}{H}}. \label{gamma2}
\end{equation}
Comparing the tunneling rate with the Boltzmann factor,
we find the temperature of the
bubble, which is given by
\begin{equation}
T_{H}=\frac{H}{2\pi}. \label{t}
\end{equation}

In principle, the inflationary universe should radiate all particles in
the standard model with a black body spectrum. To obtain the radiation
rate of fermionic particles across the horizon of de Sitter spacetime,
one should replace the Klein$-$Gordon equation with the Dirac
equation in curved spacetime, and the similar results as those in Eq.~(\ref{t})
can be obtained~\cite{lgr11,km08}.

\subsection{Reheating}

The Hubble constant of the present universe is $H_{now}\approx2.29\times
10^{-18}\operatorname{s}^{-1}$,
so the temperature $T$ at present is about
\begin{equation}
T=\frac{\hbar H_{now}}{2\pi k_{B}}\approx2.78\times10^{-30}%
\operatorname{K}.
\end{equation}
This temperature is much less than the temperature of microwave background
radiation. Except the inflation period, the universe has a very small Hubble
parameter $H\ll1/t_{p}$, so the effect of particle creation is negligible
after the inflation exits.

Let us estimate whether Hawking radiations created by inflation
have the capacity of reheating the universe or not.
In the minisuperspace model, all observers would see
a horizon at $\widetilde{r}_{H}$ and a Hawking temperature $T_{H}$. So, the
temperature of the universe is identical everywhere during inflation. Because
the Hubble parameter changes slowly, according to the Stephan$-$Boltzmann
radiation law, the energy density $\rho_{end}$ at the end of inflationary
universe is
\begin{equation}
\rho_{end}=\frac{4\sigma}{c}T_{H}^{4}=\frac{\hbar H^{4}}{240\pi^{2}c^{3}%
}.\label{rou end}%
\end{equation}
Here, $\sigma=\pi^{2}k_{B}^{4}/60\hbar^{3}c^{2}$ is the Stephan$-$Boltzmann
constant. The universe was dominated by radiations after inflation, and then
it graded into the stage dominated by matter.
Suppose that it was at $t_{m}$ that matter began to dominate the
universe with energy density $\rho_{m}$.
In the radiation dominance stage, the energy density $\rho\propto a^{-4}$ and
the scalar factor $a\propto t^{1/2}$, so we have%
\begin{equation}
\frac{\rho_{end}}{\rho_{m}}=\left(  \frac{a_{m}}{a_{end}}\right)  ^{4}=\left(
\frac{t_{m}}{t_{end}}\right)^{2}. \label{rou radiation}%
\end{equation}
In matter dominance phase, the energy density $\rho\propto a^{-3}$ and the
scalar factor $a\propto t^{2/3}$, so we have%
\begin{equation}
\frac{\rho_{m}}{\rho_{now}}=\left(  \frac{a_{now}}{a_{m}}\right)  ^{3}=\left(
\frac{t_{now}}{t_{m}}\right)^{2}. \label{rou matter}%
\end{equation}
Combine the Eqs.~(\ref{rou radiation}) and (\ref{rou matter}), we can obtain%
\[
\frac{\rho_{end}}{\rho_{now}}=\frac{\rho_{end}}{\rho_{m}}\frac{\rho_{m}}%
{\rho_{now}}=\left(  \frac{t_{now}}{t_{end}}\right)  ^{2}.%
\]

According to the Planck data, the age of the universe is 13.82 billion years,
that is, $t_{now}\approx4.36\times10^{17}\operatorname{s}$, and the matter density
(including dark matter and ordinary matter)
of the present universe is
$\rho_{now}\approx\Omega_{m}\rho_{c}\approx2.6\times10^{-27}\operatorname{kg}/
\operatorname{m}^{3}$. From equation (\ref{e:folding:number}), we can get
$t_{end}=N/H$. In the standard cosmological model, $N\approx60$.
In our model, it is sound that the Hubble parameter $H$ takes the value
$H\sim1/t_{p}\sim1.8\times10^{43}\operatorname{s}^{-1}$.
From Eq. (\ref{rou end}), we can get
\[
\rho_{end}\left(  \frac{t_{end}}{t_{now}}\right) ^{2}=1.21\times10^{-25}%
\operatorname{kg}/\operatorname{m}^{3}.
\]
Then, we have $\rho_{end}\left(  \frac{t_{end}}{t_{now}}\right)  ^{2}%
>\rho_{now}$, which suggests that the particles created by inflation have the
capability of reheating the universe and of being the source of the matter in
the universe.

The origin of the matter$-$antimatter asymmetry is one of the great questions in
cosmology. According to tunneling picture, particle and anti-particle should
be created at the same time and with the same quantity. However, there is good
evidence that there are no large regions of antimatter at any but cosmic
distance scales~\cite{cas98,mar04}. It was Sakharov who first suggested that
the baryon density might not represent some sort of initial condition, but
might be understandable in terms of microphysical laws~\cite{sad67}. He listed
three ingredients to such an understanding: (1) baryon-number violation, (2)
charge parity violation, and (3) departure from thermal equilibrium. If the
reheating temperature is greater than the mass of the gauge bosons, one can
generate the observed baryon asymmetry by charge parity violation decays of these bosons.
Baryon asymmetry can also be generated by the decay of Higgs bosons if the
reheating temperature is at least $10^{11}\operatorname{GeV} $~\cite{mar04}.
In our calculation, the temperature is high enough ($T_{H}\sim2\times
10^{31}\operatorname{K}$, or the energy scale is about
$2\times10^{18}\operatorname{GeV}$) to satisfy
these conditions, so it is possible to generate baryon asymmetry during the
inflationary universe.

\section{Energy particle conversion}

The temperature of inflationary universe can also be obtained by the Unruh
effect. The Unruh temperature, derived by William Unruh in 1976~\cite{un76},
is the effective temperature experienced by a uniformly accelerating detector
in a vacuum field:
\begin{equation}
T=\frac{\kappa}{2\pi}, \label{ut}
\end{equation}
where $\kappa$ is acceleration. Inserting the surface gravity on horizon into
Eq.~(\ref{ut}), one can easily recover the result in Eq.~(\ref{t}).

For flat FRW metric, the gravity at $\widetilde{r}$ is~\cite{rs05}
\begin{equation}
\kappa=-\frac{\widetilde{r}}{2}\left(  \dot{H}+2H^{2}\right)  .
\end{equation}
Since the horizon is located at $\widetilde{r}=1/H$, we can get $\kappa=-H$.
Here, the minus sign indicates that the direction of radiation flux
from the cosmological
horizon is opposite to the radiation flux from a black hole horizon.
For black holes, the positive energy particles escape from the
event horizon to asymptotic infinity. However, for the horizon of FRW
spacetime, the positive energy particles go inward from the horizon.
The temperature is
$T=\mid\kappa\mid/2\pi=H/2\pi$ for an empty de Sitter spacetime.

Once there is a particle created by inflation, the spacetime is no longer
empty. In this case, the gravity of the particle will affect the cosmological
horizon and the surface gravity.
When there is a particle with positive energy
$\omega$ in de Sitter spacetime, the Einstein field equation reads
\begin{align}
& \left(  \frac{\dot{a}}{a}\right)  ^{2}    =\frac{8\pi G}{3}%
\rho_{\gamma}+\frac{\Lambda}{3},\label{Einstein1}\\
& \frac{\ddot{a}}{a}   =-\frac{4\pi G}{3}\left(\rho_{\gamma}
+3p_{\gamma}\right)  +\frac{\Lambda}{3}. \label{Einstein2}
\end{align}
Here, the energy $\omega$ is relativistic particle, so we have $p_{\gamma}
=\rho_{\gamma}/3$, and $\rho_{\gamma}=3\omega H_{f}^{3}/4\pi\ll\Lambda/3$.
$H_{i}$ represents the Hubble constant for an empty de Sitter spacetime, and
$H_{f}$ represents the Hubble constant after the energy $\omega$ emit into de
Sitter spacetime. $H_{i}$ can be obtained by setting $\rho_{\gamma}=0$ in
Eq.~(\ref{Einstein1}) as $H_{i}=\sqrt{\Lambda/3}.$ Using Eqs.
(\ref{Einstein1}) and (\ref{Einstein2}), we can get
\begin{align}
\dot{H_{f}}  &  =-\frac{16\pi G}{3}\rho_{\gamma}= -4G\omega
H_{f}^{3},\label{dh/dt}\\
H_{f}  &  =\sqrt{\frac{8\pi G}{3}\rho_{\gamma}+H_{i}^{2}} =\sqrt{2G\omega
H_{f}^{3}+H_{i}^{2}}.
\label{hf}
\end{align}
When there is a particle $\omega$, the surface gravity on the horizon becomes
\begin{equation}
\kappa_{f}=-\frac{1}{2H_{f}}\left(  \dot{H_{f}}+2H_{f}^{2}\right)  .
\label{kappaf}%
\end{equation}
Inserting (\ref{dh/dt}) and (\ref{hf}) into (\ref{kappaf}), and
expanding $\kappa$ in power of $\omega$, we can get
\begin{align}
\mid\kappa_{f} \mid &  =\frac{H_{i}^{2}}{H_{f}}\approx\frac{H_{i}}%
{\sqrt{2G\omega H_{i}+1}}\nonumber\\
&  \approx H_{i}(1-G\omega H_{i}). \label{kappaf2}%
\end{align}
Here, we have used the fact that the energy of the tunneled particle $\omega$ is small,
and $H_{i}$ and $H_{f}$\ only have a tiny difference.
The temperature of the universe can thus be obtained as
\begin{equation}
T_{f}=\frac{H_{i}(1-G\omega H_{i})}{2\pi}, \label{tf}%
\end{equation}
which is lower than the temperature of the empty de Sitter spacetime $T_{i}$.
This result is physically reasonable because it is consistent with
(\ref{t}) when $\omega \rightarrow 0$.
The expression (\ref{dh/dt}) shows that
$\dot{H_{f}}<0$, which means the Hubble constant will decrease and the
horizon radius expends after the energy $\omega$ enters the cosmological horizon.
At the beginning of the particle creation, there were a few
particles in space, so the Hubble parameter decrease very slowly. With the time
increasing, there are more and more particles created in the space, which induce
a rapid decrease of the Hubble parameter and the temperature.
In this way, the energy of quantum potential changes to particles by exponential
expansion of the space and the inflation turns off.

\section{Discussion and conclusion}

For inflation driven by a scalar inflaton field $\phi$,
the power spectra of curvature and tensor perturbations are defined as~\cite{bt06}
\begin{align}
\mathcal{P}_{R} &  \simeq\left(  \frac{V^{3}}{m_{p}^{6}(V^{\prime})^{2}
}\right)  _{k=aH},\label{pr}\\
\mathcal{P}_{T} &  \simeq\left(  \frac{V}{m_{p}^{4}}\right)  _{k=aH}, \label{pt}
\end{align}
where $V(\phi)$ is the potential of the scalar inflaton field $\phi$ and the
primed means derivative with respect to $\phi$. For the slow-roll inflation
model, we have the energy density of the scalar field $\rho=\dot{\phi}%
^{2}/2+V(\phi)\approx V(\phi)$, and $\rho=3m_{p}^{2}H^{2}/8\pi$. The scalar
and tensor perturbations are related by $\mathcal{P}_{T}=r\mathcal{P}_{R}$.
Observations~\cite{par14} tell us that $r\approx0.2$ and $\mathcal{P}_{R}\approx10^{-9}$,
which yield $\mathcal{P}_{T}\approx10^{-10}$, $\rho\approx10^{-10}m_{p}^{4}$,
and $H\approx10^{-5}m_{p}$.
We should point out that the result $H\lesssim10^{-5}m_{p}$ is model dependent.
In our inflation model, there is no scalar inflaton field.
It is not obvious whether these results can be
directly applied to our model or not. So, in our calculation of reheating,
$H$ takes a value at Planck scale.
How to determine the value of Hubble parameter will be studied in future.

In summary, we have discussed the expansion solutions of a small true vacuum bubble.
We found there is an extra term called quantum potential in the Hamilton$-$Jocabi equation
after the action of the bubble was quantized. The exponential expansion solutions of
the bubble can be obtained with a specific operator ordering $p=-2$ (or 4).
Numerical calculations show that the Hubble parameter $H=\dot{a}/a\sim 1/t_p$
during the inflationary stage,
and the exponential expansion will end when the scale factor approaches $a\sim l_p$.
The value of quantum potential plus classical potential is proportional to $a^4$ for a small
bubble ($a\lesssim l_p$), while it decreases rapidly after the bubble grows up ($a>l_p $).
This indicates that the quantum potential of the vacuum bubble satisfies the conditions
required by slow-roll inflation. Thus, we can conclude that it is the quantum potential
of the vacuum bubble that plays the role of the scalar field potential assumed in the
slow-roll inflation model.

We have also studied particle creation by inflation with the picture of quantum tunneling
through the cosmological horizon.
We show that the particle production mechanism is similar to Hawking radiation of a black hole
with time inverse. The temperature at the end of inflation is
$T_H \sim 10^{31}\operatorname{K} \approx 10^{18} \operatorname{GeV}$, which suggests that particles created
by inflation have the capability of reheating the universe and being the source of the matter
in the universe.

\section*{Acknowledgement}

This work is supported by the NSFC under Grant No. 61471356.

\end{document}